\documentclass{iau}
\usepackage{graphicx}

\begin{document}

\title[Magnetic fields in single late-type giants in the Solar vicinity.] 
{Magnetic fields in single late-type giants \\ in the Solar vicinity: How common is magnetic activity on the giant branches?}

\author[Renada Konstantinova-Antova et al.]   
{Renada Konstantinova-Antova$^{1,2}$, 
 Michel Auri{\`e}re$^2$,
 Corinne Charbonnel $^{3,2}$,
 Natalia Drake $^{4,5}$,
 Gregg Wade $^6$,
 Svetla Tsvetkova $^1$,
 Pascal Petit $^2$,
 Klaus-Peter Schr{\"o}der $^7$
 \and Agnes L{\`e}bre $^8$}

\affiliation{$^1$Institute of Astronomy and NAO, BAS \\ email: {\tt renada@astro.bas.bg} \\[\affilskip]
$^2$IRAP, UMR 5277, CNRS and Universit{\`e} Paul Sabatier, Toulouse, France \\[\affilskip]
$^3$Geneva Observatory, University of Geneva, Switzerland \\[\affilskip]
$^4$Sobolev Astronomical Institute, St. Petersburg State University, St. Petersburg, Russia \\[\affilskip]
$^5$Observatorio Nacional/MCTI, Rio de Janeiro, Brazil \\[\affilskip]
$^6$Department of Physics, Royal Military College of Canada, Kingston, Ontario, Canada \\[\affilskip]
$^7$Departamento de Astronomia, Universidad de Guanajuato, Guanajuato, Mexico \\[\affilskip]
$^8$LUPM - UMR 5299 - Universit{\'e} Montpellier II/CNRS, 34095, Montpellier, France}

\pubyear{2014}
\volume{302}  
\pagerange{119--126}
\setcounter{page}{1}
\jname{Magnetic Fields Throughout Stellar Evolution}
\editors{A.C. Editor, B.D. Editor \& C.E. Editor, eds.}

\maketitle

\begin{abstract}
We present our first results on a new sample containing all single G,K and M giants down to V = 4 mag in the Solar vicinity, suitable for
spectropolarimetric (Stokes V) observations with Narval at TBL, France. For detection and measurement of the magnetic field (MF), the Least Squares Deconvolution (LSD) method was applied (\cite[Donati et al. 1997]{Donati_etal97}) that in the present case enables detection of large-scale MFs even weaker than the solar one (the typical precision of our longitudinal MF measurements is 0.1-0.2 G). The evolutionary status of the stars is determined on the basis of the evolutionary models with rotation (\cite[Lagarde et al. 2012]{Lagarde_etal12}; Charbonnel et al., in prep.) and fundamental parameters given by \cite[Massarotti et al. (1998)]{Massarotti_etal08}. The stars appear to be in the mass range 1-4 $M_{\odot}$, situated at different evolutionary stages after the Main Sequence (MS), up to the Asymptotic Giant Branch (AGB). 

The sample contains 45 stars. Up to now, 29 stars are observed (that is about 64 $\%$ of the sample), each observed at least twice. For 2 stars in the Hertzsprung gap, one is definitely  Zeeman detected. Only 5 G and K giants, situated mainly at the base of the Red Giant Branch (RGB) and in the He-burning phase are detected. Surprisingly, a lot of stars ascending towards the RGB tip and in early AGB phase are detected (8 of 13 observed stars). For all Zeeman detected stars $v\,\sin\,i$ is redetermined and appears in the interval 2-3 km/s, but few giants  with MF possess larger $v\,\sin\,i$.
\end{abstract}

\firstsection 
\section{Introduction}

Single late-type giants are an excellent laboratory to study the conditions under which dynamo could operate at different stages of the stellar evolution when significant changes in the structure of intermediate mass stars appear. Different hypotheses were suggested to explain the origin of the magnetic field and activity in giant stars: dynamo operation as a result of planet engulfment (\cite[Siess \& Livio 1999]{SiessLivio99}) or angular momentum dredge-up from the interior (\cite[Simon \& Drake 1989]{SimonDrake89}) for the faster rotators; remnant rotation for the more massive stars; Ap star descendants for slow rotators with a strong MF (\cite[Stepien 1993]{Stepien93}). 
Our study tries to give an answer what kind of dynamo that operates in RGB and AGB stars, what are the properties of the magnetic activity in single giants, how long activity lasts and how common is it. 

Our first sample of single G and K giants (more than 50 stars) revealed that about 50 $\%$ of these stars possess MF (Auri{\`e}re et al., in prep.; \cite[Konstantinova-Antova et al. 2013]{Konst-Antova_etal13}). The stars are situated mostly in the first dredge-up phase and some are in the He-burning phase. Magnetic field was detected also in stars ascending the RGB and AGB single giants (\cite[Konstantinova-Antova et al. 2010]{Konst-Antova_etal10}; \cite[Konstantinova-Antova et al. 2013]{Konst-Antova_etal13}). 
However, this sample was biased because of its selection was made on the basis of fast rotation and/or activity signatures. 

\section{Selection of the sample. Equipment and Methods.}

To understand to what extent late-type giants possess MFs, a new sample entirely independent of activity was selected. Our new sample contains all single G, K and M giants
up to $V=4$ mag in the Solar vicinity, accessible for
observations at Pic du Midi with Narval at TBL. These are 45 stars. Their $v\,\sin\,i$, log(L) and $T_{eff}$ are specified by \cite[Massarotti et al. (1998)]{Massarotti_etal08}. The stars cover the different evolutionary stages
after the MS:
Hertzsprung gap (5 stars), the region of the 
dredge-up phase and He-burning
(21 stars), stars ascending the RGB and AGB
(19 stars). Their observations enable a precision of Bl measurement of typically
0.2 G, using the LSD method (\cite[Donati et al. 1997]{Donati_etal97}). The stars are observed at least
twice with a time interval between the observations of one month and more. For a few stars we also used data obtained in the period 2008 -2010 with Narval at TBL and its twin ESPaDOnS at CFHT.

\section{First results}

29 of 45 stars are observed up to now (64 $\%$ of the sample).
15 of the observed stars (52 $\%$) are Zeemen detected. For the different evolutionary stages: 
Hertzsprung gap - 2 stars of 5 observed. 40 $\%$
of the stars are observed, 50 $\%$ of the observed
stars detected.
Base of the RGB - 14 of 21 observed. 67 $\%$
observed, 35 $\%$ of the observed stars are detected.
Stars ascending towards the RGB tip and AGB stars - 13 of 19 stars observed. 68 $\%$
observed, 61 $\%$ of the observed stars are detected.

\begin{figure}
\begin{center}
 \includegraphics[width=2.4in,height=1.9in]{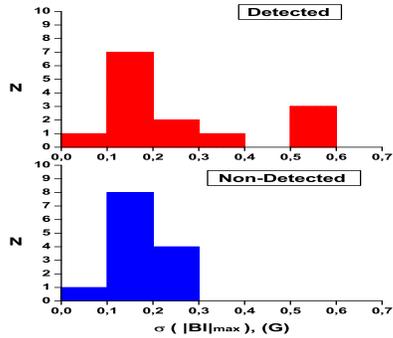} 
 \caption{One $\sigma$ error bars of the $B_l$ measurements for the Zeeman detected and non-detected stars.}
   \label{fig1}
\end{center}
\end{figure}

The precision of the $B_l$ determination (Fig.\,\ref{fig1}) indicates that our aim for a deep magnetic survey is fulfilled. The distribution of error bars confirms that the non-detections are generally not due to poor quality observations.

\begin{figure}
\begin{center}
\includegraphics[width=2.6in,height=2.1in]{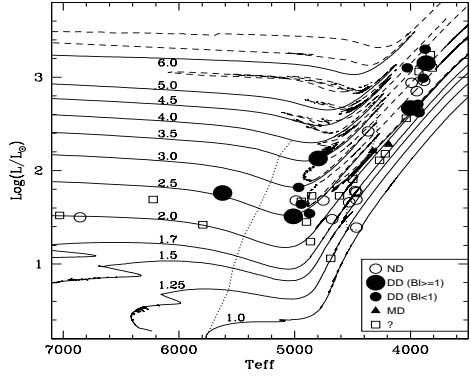} 
 \caption{Situation of the studied stars on the HR diagram. Filled circles stand for detections, open ones - for non-detections, triangles - for marginal  detections. Squares indicate the stars not observed yet. The dotted lines delimit the first dredge-up phase. Dashed lines designate the He-burning and AGB phases.}
   \label{fig2}
\end{center}
\end{figure}

The situation of our sample stars on the HRD is presented in Fig.\,\ref{fig2}. Models with rotation for solar chemical composition (\cite[Lagarde et al. 2012]{Lagarde_etal12}; Charbonnel et al., in prep.) are used. The detected stars are in the interval 1.7-4 $M_{\odot}$. There is a tendency for "clumping" at certain evolutionary phases: the first dredge-up and He-burning region, and the region near the upper RGB and early AGB for less massive giants, and the AGB for the more massive ones. Stars with weak and stronger MFs occupy the same regions of the HRD. Only two stars with very weak MF (and marginal detection) appear in the region between the first dredge-up and the tip of the RGB. Also detected and non-detected stars coexist in one and the same regions in the HRD.

\begin{figure}
\begin{center}
 \includegraphics[width=2.6in, height=1.9in]{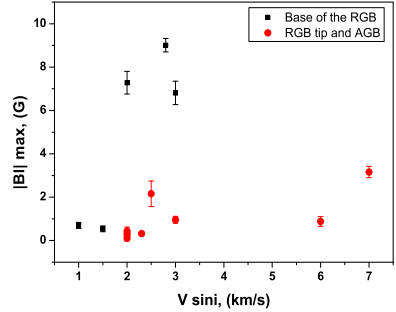} 
 \caption{Dependence of $|B_l|_{max}$ on $v\,\sin\,i$ for the Zeeman detected giants. Squares stand for the region of the base of RGB and He-burning phases, and circles stand for the stars ascending towards the RGB tip and AGB stars.}
   \label{fig3}
\end{center}
\end{figure}

According to \cite[Gray (2013)]{Gray13}, it appeared that the $v\,\sin\,i$ values of \cite[Massarotti et al. (1998)]{Massarotti_etal08} could be overestimated. We performed a redetermination of $v\,\sin\,i$ for the detected stars, using the spectrum
synthesis method and taking into account the macroturbulence. Most of them have $v\,\sin\,i$ in the interval 2-3 km/s. The dependence of $|B_l|_{max}$ on $v\,\sin\,i$ is also presented (Fig.\,\ref{fig3}). There are some indications that stars with higher $v\,\sin\,i$ situated in the second "magnetic" region in the HRD possess stronger MF. However, further study is necessary on this topic, including more $B_l$ measurements and determination of the rotation period.

Recent findings by \cite[Mosser et al. (2012)]{Mosser_etal12} that core rotation slows down during the RGB phase support the idea that angular momentum dredge-up could be essential for the dynamo operation in most magnetic single giants. Further studies on the processes of angular momentum transfer between the core region and the convective envelope in giants are required.  


\section{Conclusions}

The X-ray study by \cite[Schroder et al. (1998)]{Schroder_etal98} gave some hints that all giants in
the Solar vicinity could be active at least at the solar level, but the first results of our
deep magnetic study do not confirm this expectation (52 $\%$ of the stars
observed up to now possess MF within the limit of our precision).
We study a complete sample of single G,K and M giants in the solar vicinity, selected
independently of activity signatures, rotation, etc. The mass range appears to be in the interval 1-4
$M_{\odot}$. The sample has better coverage of the different evolutionary stages as compared to our
previous one. The studied stars appear to have small $v\,\sin\,i$, except few
M giants. Stars with weak and stronger magnetic field coexist in 
the same place on the HRD. We find no evidence for magnetic giants below 1.7 $M_{\odot}$.
The magnetic G and K giants are at the first dredge-up and He-burning phases.
A lot of stars with MF detection are located at the upper part of RGB and in the AGB, contrary to the
expectations. There are some indications for a dependence of $|B_l|_{max}$ on $v\,\sin\,i$. Is the $\alpha$ -
$\omega$ dynamo in operation there (predictions by \cite[Nordhaus et al. 2008]{Nordhaus_etal08})? For
our previous sample of 9 giants (for 7 of them MF was detected) we have a
good correlation (\cite[Konstantinova-Antova et al. 2013]{Konst-Antova_etal13}) to suspect this.
Further study of their MF variability, based on the rotation period is required to have an idea about
the dynamo type, but for the moment the periods are unknown. Future works on angular momentum transport in the giants interior are also required.

\section{Acknowledgements}

\small{We thank the TBL and CFHT teams for the service observing.
The NARVAL observations in 2008 are granted under OPTICON programs. We also acknowledge NARVAL
observations under French PNPS program and
CFHT observations under a Canadian program.
R.K.-A. is thankful for the
possibility to work for two months in 2013 in IRAP, Tarbes as invited
researcher. The work on the topic was partly supported under the
program RILA/CAMPUS. N.A.D. thanks PCI/MCTI (Brazil)
grant under the project 302350/2013-6. C.C. acknowledges support from the Swiss National Foundation. G.A.W. is 
supported by an NSERC discovery grant.
R.K.-A. and S.Ts. are thankful to the TBL for the partial financial
support to attend IAUS302.}

\end{document}